\documentclass[english,intlimits]{elsart}
\usepackage[T1]{fontenc}
\usepackage[latin1]{inputenc}
\usepackage{amsmath}
\usepackage{babel}
\usepackage{graphics}
\usepackage{amssymb}

\begin{document}

\begin{frontmatter}

\title{L\'{e}vy Flights in External Force Fields: From Models to Equations}

\author[MPISF]{D. Brockmann} and
\author[HUBERLIN]{I. M. Sokolov\corauthref{cor}}
\ead{igor.sokolov@physik.hu-berlin.de}
\corauth[cor]{Corresponding author}

\address[MPISF]{Max-Planck-Institut für Strömungsforschung, Göttingen, Germany}
\address[HUBERLIN]{Institut f\"{u}r Physik, Humboldt-Universit\"{a}t zu Berlin,  Invalidenstraße 110, D-10115 Berlin, Germany}

\begin{abstract}
We consider different generalizations of the Fokker-Planck-equation
devised to describe L\'{e}vy processes in potential force fields.
We show that such
generalizations can proceed along different lines. On one hand,
 L\'{e}vy
statistics can emerge from the fractal temporal nature of the
underlying process, i.e. a high variability in the rate of microscopic events. On the other hand, they may be a direct
consequence of the scale-free spatial
 structure on which the process evolves.
Although both forms considered lead to Boltzmann equilibrium,
the relaxation patterns are quite different.
 As an example, generalized diffusion
in a double-well potential is considered. 
\end{abstract}

\end{frontmatter}

\section{Introduction}

Random walk processes leading to anomalous diffusion  are
adequate for describing various physical situations. The continuous time
random walk (CTRW) model of Scher and Montroll~\cite{ScheM} for instance, leading
to strongly sub-diffusive behavior, was a milestone in understanding
photoconductivity in strongly disordered and glassy semiconductors.
Due to its simplicity, this model was recently employed to investigate
aging phenomena typical for glasses
and related complex systems~\cite{Montus,Maas,Aging1,Aging2}.
On the other hand, L\'{e}vy-flight models~\cite{KlaShleZu},
leading to superdiffusion, are adequate for the description of phenomena ranging from
transport in heterogeneous catalysis~\cite{Bych}, self-diffusion
in micelle systems~\cite{OBLU}, reactions and transport in polymer
systems under conformational motion~\cite{Sok0}, transport processes
in heterogeneous rocks~\cite{KBZS} and the behavior
of dynamical systems~\cite{Sporad} to flight paths of albatrosses~\cite{albi} and even
human eye-movements~\cite{brocki}. Closely related models appear
in the description of economic time series~\cite{Trunk}.  L\'{e}vy-related
statistics were observed in hydrodynamic transport~\cite{Swinney},
and in the motion of gold nanoclusters on graphite~\cite{Clusters}.
In addition, 
mixed models were proposed in which slow temporal evolution
(described by Scher-Montroll CTRW) is combined with the possibility
of L\'{e}vy-jumps, so that in general both sub- or superdiffusive
behavior can arise~\cite{Fogedby1,ralf}. 
Similarily, processes often referred to as L\'{e}vy-walks are based
on long-tailed jumps in combinations with a time cost to perform them, see
for example~\cite{KlaShleZu,shles}. L\'{e}vy-walks were introduced
in~\cite{theo1} to explain accelerated diffusion in Josephson junctions
and related chaotic systems~\cite{theo2}.

Typically, the corresponding
processes are described on a stochastic level, such that the incorporation
of boundary conditions, time- or position-dependent forces leads
to considerable mathematical difficulties.

Concerning ordinary diffusion and transport in a stochastic
system, the Fokker-Planck equation (FPE) is
an adequate instrument for solving boundary value or
time-dependent problems~\cite{gardiner}. The FPE is an equation for the probability
density function (pdf) \( p(x,t) \) of the particle's position at
time \( t\,  \)and reads \begin{equation}
\label{FPE}
\frac{\partial p(x,t)}{\partial t}=\mathcal{L}_{FP}\,p=-\nabla \left( \mu f(x,t)p(x,t)\right) +\kappa \Delta p(x,t).
\end{equation}
 Here \( \mathcal{L}_{FP} \) denotes the Fokker-Planck operator,
\( \mu  \) and \( \kappa  \) are the mobility and the diffusion
coefficient (assumed to be constant) and \( f(x,t) \) is the external force
acting on the particle. Note that since the mobility and the diffusion coefficient
are connected via the Einstein relation, \( \kappa =kT\mu \equiv \mu /\beta  \),
the equation effectively contains only one free parameter.
Alternatively, eq.~(\ref{FPE}) may be recast into a different form 
in the case of a potential force \( f(x,t)=-\nabla \Phi (x,t) \): 
\begin{equation}
\label{FPEvar}
\frac{\partial p(x,t)}{\partial t}=\kappa\left[ e^{-\beta \Phi /2}\left( \frac{\partial }{\partial x}\right) ^{2}e^{\beta \Phi /2}\, p-p\, e^{\beta \Phi /2}\left( \frac{\partial }{\partial x}\right) ^{2}e^{-\beta \Phi /2}\right] ,
\end{equation}
 the equivalence of which to eq.~(\ref{FPE}) is easily checked by expanding
the differentiations. Obviously, the stationary solution of the FPE
 (if it exists) is the Boltzmann distribution, \( p(x)\propto \exp (-\beta \Phi ) \). 
Inserting this form into eq.~(\ref{FPEvar})
one readily infers that the two terms entering
with opposite signs are equal and cancel.

In order to investigate analogous systems incorporating  anomalous diffusion, various
generalizations of the FPE were proposed.
Concerning subdiffusion the appropriate generalizations can be obtained
from CTRW models, for a general survey see~\cite{metzler}, as well as
from an approach based on fractional master equations~\cite{hilfer1}.
 Concerning superdiffusion
(L\'{e}vy-flights), such generalizations typically involve changing
from a gradient and Laplacian to the corresponding fractional derivative.
For example, a pure L\'{e}vy-flight can be described by means of the
fractional diffusion equation~\cite{sesha,peseckis,foggy2} 
\begin{equation}
\label{D1}
\frac{\partial p(x,t)}{\partial t}=\kappa_{\alpha} \Delta ^{\alpha /2}p(x,t),
\end{equation}
 where \( \Delta ^{\alpha /2} \) is a fractional Laplacian, a linear
operator, whose action on a function \( f(x) \) in Fourier space
is described by \( \Delta ^{\alpha /2}f(x)=-\left( k^{2}\right) ^{\alpha /2}f(k)=- \left| k\right| ^{\alpha }f(k) \).
The coordinate representation of this operator is discussed in 
Appendix 1. The solution of eq.~(\ref{D1})
in Fourier space reads: \begin{equation}
G(k,t)=\exp (-\kappa \left| k\right| ^{\alpha }t)
\end{equation}
 which in coordinate space corresponds to a symmetric L\'{e}vy stable
distribution \begin{equation}
G(x,t)=\frac{1}{(\kappa t)^{1/\alpha }}L\left( \frac{x}{(\kappa t)^{1/\alpha }};\alpha ,0\right) 
\end{equation}
 (we use here the canonical notation, see ref.~\cite{Feller}). On
the other hand, there is no generic way to
generalize the Fokker-Planck operator containing an external force term:
For example,  the drift term may stay unchanged, may correspond
to a symmetric or asymmetric fractional derivative, or the whole Fokker-Planck-operator 
may be raised to the power of \( \alpha /2 \). All corresponding
equations have their physical meaning and applications. The properties
of their solutions differ considerably. Moreover,
as we proceed to show, other generalizations are also possible. The
situation here is to some extent similar to one in fractal geometry: 
What is the generalization of the Euclidean dimension to
the fractal case? Depending on the system and on the property of interest
it could be the fractal dimension, the spectral dimension, or even
a spectrum of different dimensions. In what follows we discuss the
properties of some generalization of the FPE to the superdiffusive
case leading to  L\'{e}vy-flights in the force-free limit. We will focus on two
complementary situations, namely one in which superdiffusion stems 
from fractal temporal properties of the underlying microscopic dynamics and 
another in which the spatial structure is responsible for enhanced
diffusion. We mostly concentrate on the generalizations of the forms,
eq.~(\ref{FPE}) and eq.~(\ref{FPEvar}) which lead to a Boltzmann distribution
in equilibrium. We discuss the physical implications of the corresponding
equations and some properties of their solutions. As a physically
most interesting application, a fractional generalization of a Kramers
problem in a double-well potential is considered. Let us first turn
to the situations which can be considered as stemming from the fractal
temporal behavior.

\section{L\'{e}vy processes as stemming from subordination}

An important property of symmetric
L\'{e}vy processes is the fact that they are subordinated
to ordinary Brownian motion: the corresponding pdf can be represented
in the form:
\begin{equation}
\label{Subord}
p(x,t)=\frac{1}{t^{1/\alpha }}L\left( \frac{x}{t^{1/\alpha }};\alpha ,0\right) =\int _{0}^{\infty }\frac{1}{\sqrt{2\pi \tau }}\exp \left( -\frac{x^{2}}{2\tau }\right) \frac{1}{t^{2/\alpha }}L\left( \frac{\tau }{t^{2/\alpha }};\frac{\alpha }{2},1\right) d\tau 
\end{equation}
 where \( L\left( x;\alpha /2,1\right)  \) is an extreme (one-sided)
L\'{e}vy distribution of index \( \alpha /2 \)~\cite{Feller}. 

The
variable \( \tau  \) is called the operational time of the process.
The interpretation of eq.~(\ref{Subord}) is that Levy-flights can
be considered as stemming from a highly irregular sampling  trajectories
generated by simple diffusion (random walk). 
This type of temporal behavior plays a very important role
in many situations encountered in physics, such as anequilibrium phase
transitions~\cite{hilfer2}. For a review see~\cite{hilfer3} 
and references therein.

 Here, the trajectory of diffusion
(a random walk in a discrete case or a Wiener path in a continuous
situation) is parameterized by the operational time \( \tau  \) (say,
the number of steps of the random walk), which itself is a random
function of the (physical) time. The random process \( \tau (t) \)
is a process with positive increments, and the distribution of \( \tau (t) \)
is given by a one-sided L\'{e}vy law. Thus, the increase \( \Delta \tau  \)
of the operational time per physical time unit is given by a one-sided
L\'{e}vy-distribution, \( p(\Delta \tau )=L\left( \Delta \tau ;\alpha /2,1\right)  \)
having a power-law tail, \( p(\Delta \tau )\propto \tau ^{-1-\alpha /2} \).
\begin{figure}
{\centering \resizebox*{12cm}{!}{\includegraphics{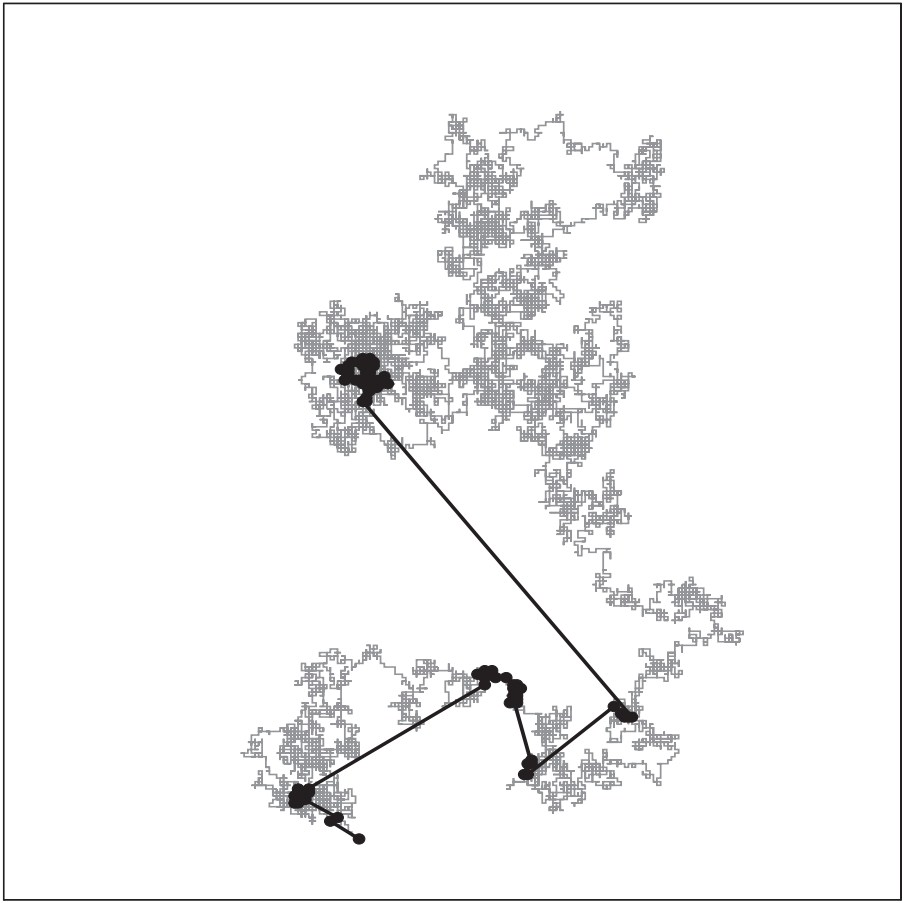}} \par}

\caption{\label{fig:subord} The mechanism of subordination:
L\'{e}vy-flights can be considered as stemming
from a (highly inhomogeneous) random sampling of a simple random walk.
}
\end{figure}

The corresponding situation is depicted in fig. \ref{fig:subord}. Here a L\'{e}vy-flight
of \( N=100 \) steps is generated as follows. One generates a sequence
\( \left\{ n_{i}\right\}  \) \( (i=1,2,...,N) \) of the numbers
of steps between the observation instants and plots a realization
of a simple random walk of \( M=\sum _{i=1}^{N}n_{i} \) steps. The
turning points of a L\'{e}vy-flight are then the positions of a
random walker after \( m_{k} \) steps, \( m_{k}=\sum _{i=1}^{k}n_{i} \).
Fig. \ref{fig:subord} corresponds to a process subordinated to a random walk
 under the operational
time given by 
\begin{equation}
p(n_{i})=\frac{1}{\sqrt{\pi }n_{i}^{3/2}}\exp \left( -\frac{1}{2n_{i}}\right) .
\end{equation}
 The overall pdf of the process \( p(x,t) \) is then given approximately by a Cauchy
distribution, 
\begin{eqnarray}
p(x,t)  & = & \sum _{n=0}^{\infty }\frac{1}{\sqrt{2\pi n}}\exp \left( -\frac{x^{2}}{2n}\right) \frac{t}{\sqrt{\pi }n^{3/2}}\exp \left( -\frac{t^{2}}{2n}\right)  \simeq \nonumber \\
& \simeq & \int _{0}^{\infty }\frac{1}{\sqrt{2\pi n}}\exp \left( -\frac{x^{2}}{2n}\right) \frac{t}{\sqrt{\pi }n^{3/2}}\exp \left( -\frac{t^{2}}{2n}\right) dn 
= \frac{t}{\pi (t^{2}+x^{2})}.
\end{eqnarray}
Although this continuous approximation is valid here, some care must be 
taken in the general case when sums over steps in random walks are replaced 
by integral, see for example~\cite{barkai}.
Applying the subordination procedure to the general, biased diffusive
process, one is lead to the generalization of the FPE,
in which the entire Fokker-Planck-operator is raised to the power \( \alpha /2  \):
\begin{equation}
\label{FFPESub}
\frac{\partial p}{\partial t}=-\left( -\mathcal{L}_{FP}\right) ^{\alpha /2 }p
\end{equation}
 as shown in~\cite{TD}. The generalized Fokker-Planck operator
\( \mathcal{L}_{\alpha }=-\left( -\mathcal{L}_{FP}\right) ^{\alpha /2 } \)
commutes with \( \mathcal{L}_{FP} \) and shares with it the same
set of the eigenfunctions \( \phi _{i} \). The corresponding eigenvalues
\( \Lambda _{i}\,  \)of the operator \( \mathcal{L}_{\alpha } \)
are connected with those of \( \mathcal{L}_{FP} \) via \( \Lambda _{i}=-(-\lambda _{i})^{\alpha /2 } \).
Since the eigenvalues \( \lambda _{i} \) of \( \mathcal{L}_{FP} \)
are real and nonpositive, this is true for the \( \Lambda _{i} \) as well. Eq.~(\ref{FFPESub})
follows from the subordination procedure just in the same way as a
Cauchy distribution followed from a Gaussian distribution in the example given
above. Let us consider a spectral representation of the solution of
the FPE, \begin{equation}
g(x,\tau )=\sum _{i}\phi _{i}(x)e^{-\left| \lambda _{i}\right| \tau }.
\end{equation}
 The subordination procedure applied to this solution leads to
\begin{eqnarray}
p(x,t) & = & \int g(x,\tau )\frac{1}{t^{2/\alpha }}L\left( \frac{\tau }{t^{2/\alpha }};\frac{\alpha }{2},1\right) d\tau =\label{Spectr} \\
 & = & \sum _{i}\phi _{i}(x)\int e^{-\left| \lambda _{i}\right| \tau }L\left( \frac{\tau }{t^{2/\alpha }};\frac{\alpha }{2},1\right) d\tau \nonumber \\
 & = & \sum _{i}\phi _{i}(x)e^{-\left| \lambda _{i}\right| ^{\alpha }t},\nonumber 
\end{eqnarray}
 due to a well-known property of the Laplace-transforms of one-sided
L\'{e}vy distributions~\cite{Feller}. Note that the eigenfunction
\( \phi _{0} \) corresponding to a zero eigenvalue (if any) is a
stationary distribution, which is  the same (i.e. a Boltzmann
distribution) for the normal and for the generalized process. The
spectral decomposition of solutions of eq.~(\ref{FFPESub}) given by
eq.~(\ref{Spectr}) is incorporated in our numerical procedure for
the solution of superdiffusive FPE in a potential field considered
in section \ref{compa}.

Eq.~(\ref{FFPESub}) can also be interpreted within the Langevin scheme,
which clarifies the circumstances under which Boltzmann 
statistics may or may not appear when superdiffusive 
generalizations of the FPEs
are introduced.
Let us first consider a purely diffusive situation without external
force. In the continuous limit (corresponding to averaging over
the time periods which are short enough to consider all parameters
constant but during which many steps of a random walk are performed),
the system's development in its operational time is given by a Langevin
equation \begin{equation}
\frac{d}{d\tau }x(\tau )=\sqrt{2\kappa }\xi (\tau )
\end{equation}
 where \( \xi (\tau ) \) is a \( \delta  \)-correlated Gaussian
noise with zero mean and with unit dispersion. Subordination requires
that
\begin{equation}
\frac{d\tau }{dt}=\lambda (t),
\end{equation}
 where \( \lambda (t) \) is a one-sided L\'{e}vy process of index
\( \alpha /2 \). The resulting process is thus described by a Langevin
equation \begin{equation}
\frac{dx}{dt}=\frac{d\tau }{dt}\frac{d}{d\tau }x(\tau )=\sqrt{2\kappa }\lambda (t)\xi (\tau (t)),
\end{equation}
 where \( \tau (t) \) is a monotonously increasing random function which
is changing slowly on the scale on which \( \xi (\tau )\,  \)is correlated.
This suggests that the random process \( \xi (\tau (t)) \) is a Gaussian
one, has zero mean and dispersion \( \left\langle \xi (\tau (t))\xi (\tau (t^{\prime }))\right\rangle =\delta \left( \tau (t)-\tau (t^{\prime })\right) =(d\tau /dt)^{-1}\delta (t-t^{\prime })=\lambda (t)^{-1}\delta (t-t^{\prime }) \),
indicating that the process \( x(t)\,  \)can be formally represented by a
Langevin equation \begin{equation}
\dot{x}(t)=\sqrt{2\kappa \lambda (t)}\eta (t)
\end{equation}
 where \( \eta (t) \) is a \( \delta  \)-correlated Gaussian noise
with zero mean and unit dispersion. From this equation, a FPE
of the type of eq.~(\ref{D1}) follows along the usual lines.

Physically, such strongly inhomogeneous behavior can be attributed
to random fluctuations of the mobility or of the temperature or of
the both: \( \kappa (t)=\kappa \lambda (t)=k\mu (t)T(t) \). Let us return
to the full process described by a Langevin equation \begin{equation}
\dot{x}(t)=\mu f+\sqrt{2\mu kT}\xi (t)
\end{equation}
 with a potential force \( f(x,t)=-\nabla U(x,t) \). Here the situations
corresponding to the fluctuations in temperature and to ones in mobility
differ vastly. Keeping the temperature constant and letting \( \mu  \)
fluctuate leads to a process subordinated to a biased random walk,
which is described by a Langevin equation \begin{equation}
\dot{x}(t)=\mu (t)f+\sqrt{2\mu (t)kT}\xi (t)
\end{equation}
 where \( \mu (t) \) is now a strongly fluctuating random function,
whose distribution is given by a one-sided L\'{e}vy law. The corresponding
pdf is governed exactly by a fractional Fokker-Planck equation, eq.~(\ref{FFPESub})
derived above. This equation shows a relaxation to a normal Boltzmann
distribution, which is not surprising since the corresponding scheme describes
a system showing detailed balance at given, fixed temperature \( T \). 
Note that although the equilibrium properties of the system are usual, 
its relaxation properties are not, see ref.~\cite{Levycont} for a detailed
discussion. Thus, it does not exhibit linear response to
a constant external field, and, moreover, since the corresponding pdf neither 
possesses the second moment nor scales, no reasonable generalization
of the Einstein's relation exists. The same is valid for the model of 
topological superdiffusion discussed below. On the other hand, as we 
proceed to show, the behavior of both models in confining potentials such 
as a harmonic or a double-well one, is not as exotic (vide infra). 

The situation under fluctuating temperature is vastly different: a system
with fluctuating temperature must not show Boltzmann statistics at
equilibrium (if any). Let us mention that under certain circumstances 
the distributions given by Langevin equations with fluctuating parameters
approach ones given by Tsallis statistics~\cite{Beck},
i.e. ones stemming from a thermodynamic construct in which the entropy
is non-extensive and thus the concept of temperature can not be introduced
in a usual way. 

In our case, fluctuations in temperature do not affect the first term,
i.e. the deterministic motion and lead (after performing usual steps)
to a fractional Fokker-Planck equation (FFPE) in the form: \begin{equation}
\label{Su1}
\frac{\partial p}{\partial t}=-\frac{\partial }{\partial x}\left( \mu fp\right) +\kappa \Delta ^{\alpha /2}p,
\end{equation}
whose solutions are discussed in ref.~\cite{Sune}.

\section{Topologically induced superdiffusion}

Complementary to the situation discussed in the previous section in
which superdiffusion originates from  fractal temporal behavior,
we will discuss the fractional generalization of the  FPE
adequate for the description of diffusion on scale-free structures.
Contrary to the situation investigated in the previous section in which superdiffusion
is caused by a fractal sampling of an otherwise continuous path, we
will focus on the limiting behavior of a master equation describing
a Markovian jump process in which transitions are caused by thermal
hopping between energetically different states on a topologically
complex structure, ref.~\cite{Sok0}.

\begin{figure}
{\centering \resizebox*{12cm}{!}{\includegraphics{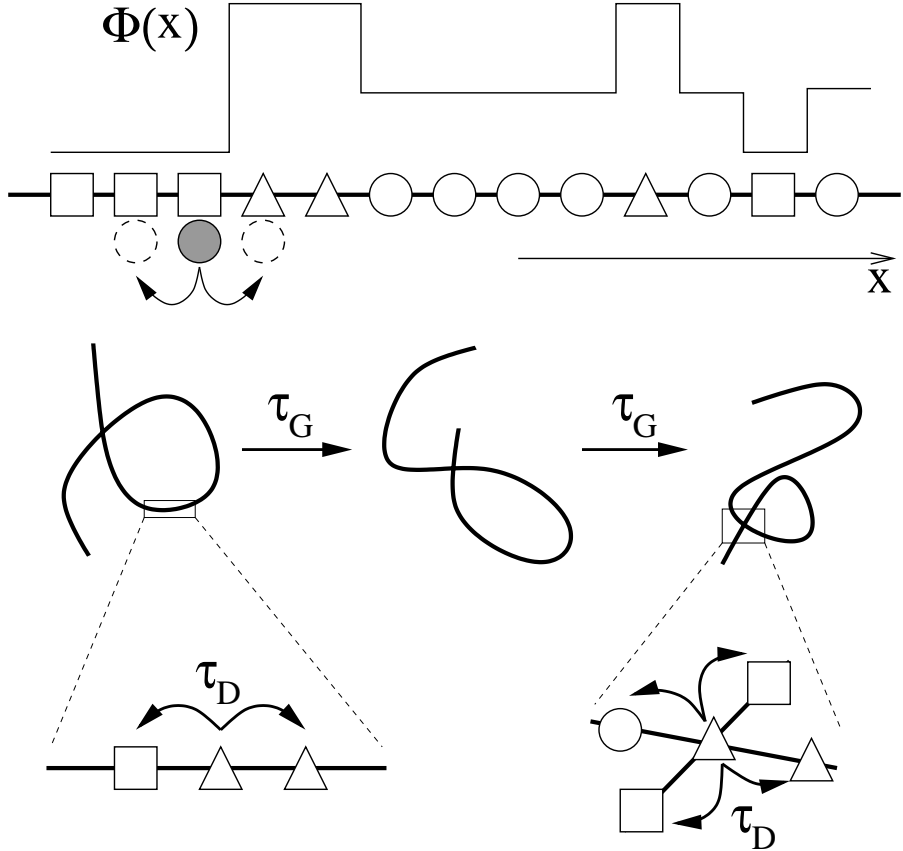}} \par}

\caption{\label{fig:polymer}Random walk of a particle on a heterogeneous polymer subjected
to fast conformational changes. The polymer consist of different types of monomers denoted by the symbols along the chain (top). Each monomer is associated with an intrinsic potential \( \Phi (x) \). 
Due to conformational changes, the walker
may jump between two sites nearby in Euclidean space
but far apart along the
chemical sequence of the polymer. }
\end{figure}
As a model system, we consider a heteropolymer in solution depicted in fig.~\ref{fig:polymer}. We focus on the dynamics of a test particle
performing thermally activated motion along the linear chemical coordinate
axis \( x \) of the polymer. If the polymer is fixed at full extend,
thermal activation causes the particle to hop at a rate \( 1/\tau _{D} \)
from a given site to a neighboring site with a probability determined
by the energy difference within a given mononer pair. However, a
polymer in solution is subjected to thermal conformational changes
occurring on a time scale \( \tau _{G} \). The conformational changes
allow sites which are far apart along the chemical sequence to come close
in Euclidian space, as indicated in figure \ref{fig:polymer}: For simplicity,
a Gaussian chain (corresponding to a Rouse dynamics) can be considered, moreover,
the conformational changes will be taken fast enough on the scale \( \tau _{D} \). 
In this case the particle
may jump from a given site \( x \) to a site \( y \) far apart with
a rate proportional to the probability that the sites \( x \) and
\( y \) have come close in the given time interval. In our case, the probability of such 
event is prortional to \( |x-y|^{-3/2} \).
As we proceed to show, the fractional generalization of the Fokker-Planck equation
corresponding to the situation depicted above differs from the ones
considered so far. 

The system introduced above can be modeled by a Markovian jump process
governed by a temporally homogeneous master equation
\begin{equation}
\label{eq:mastergleichung}
\frac{\partial p(x,t)}{\partial t}=\int dy\, \left[ w(x|y)p(y,t)-w(y|x)p(x,t)\right] ,
\end{equation}
 in which the rates \( w(x|y) \) are determined by the short time
behavior of the conditional probability to make a transition \( y\rightarrow x \)\begin{equation}
\label{eq:wdef}
w(x|y)\equiv \lim _{\Delta t\rightarrow 0}\frac{1}{\Delta t}p(x,t+\Delta t|y,t).
\end{equation}
Let us mention here, that in eq.(\ref{eq:wdef}) the rates $w(x|y)$ may be singular and may
even defy normalizability. 
The Cauchy process for example is defined by
$w(x|y)=1/\pi(x-y)^{-2}$. If, on the other hand, $\int dx w(x|y)=\text{const}$, exists
the constant can be factored out and absorbed by a rescaling of the time. In this
case the process is a pure jump process, consecutive jumps are seperated
by time periods during which the random process is constant, in contrast
to the Cauchy process realizations of which are discontinuous almost everywhere.

 In our example the coordinate \( x \) represents the chemical coordinate
along the polymer chain. The reasoning above suggests a rate of the
form \begin{equation}
\label{eq:rate}
w(x|y)=\frac{1}{\tau _{0}}e^{-\beta [\Phi (x)-\Phi (y)]/2}\, f(x-y).
\end{equation}
 The right hand side of eq.~(\ref{eq:rate}) consists of the thermal component
\( \exp (-\beta [\Phi (x)-\Phi (y)]/2) \) accounting for the fact
that transitions between energetically different states are less likely
to occur if the potential difference \( \Phi (x)-\Phi (y) \) is high.
The second factor, \( f(x-y) \) represents the probability of states
with chemical coordinates \( x \) and \( y \) to be nearby in Euclidean
space, i.e. this term accounts for the geometrical complexity of the
system. Due to the translational invariance and symmetry of the system
in Euclidean space this term is symmetric and depends on the distance
\( \left| x-y\right|  \) only. Obviously, a system defined by a rate
(\ref{eq:rate}) fulfills detailed balance, characteristic of systems
in thermal equilibrium. Note also that the stationary solution of
Eq. (\ref{eq:mastergleichung}) with the rates given by (\ref{eq:rate}) 
is the Boltzmann distribution \begin{equation}
\label{eq:boltzmann}
p_{s}(x)\propto e^{-\beta \Phi (x)}.
\end{equation}

It is instructive to consider a fully extended polymer of total length
\( L \). In this case only nearest neighbor hopping occurs, i.e.
\begin{equation}
\label{deq:myff}
f(x)=\frac{1}{2}\left( \delta (x-\sigma )+\delta (x+\sigma )\right) .
\end{equation}
 In (\ref{deq:myff}) \( \sigma \ll L \) is the spacing between adjacent
monomers. Setting \( q_{t}(x)\equiv e^{\beta \Phi (x)/2}p(x,t) \)
and \( s(x)\equiv e^{-\beta \Phi (x)/2} \) the master equation (\ref{eq:mastergleichung})
reduces to: \begin{eqnarray}
\frac{\partial p(x,t)}{\partial t} & = & \frac{1}{2\tau _{0}}s(x)\left\{ q_{t}(x+\sigma )+q_{t}(x-\sigma )\right\} \\
 &  & -\frac{1}{2\tau _{0}}q_{t}(x)\left\{ s(x+\sigma )+s(x-\sigma )\right\} \\
 & = & \frac{\sigma ^{2}}{2\tau _{0}}s(x)\left\{ \frac{q_{t}(x+\sigma )+q_{t}(x-\sigma )-2q_{t}(x)}{\sigma ^{2}}\right\} \\
 &  & -\frac{\sigma ^{2}}{2\tau _{0}}q_{t}(x)\left\{ \frac{s(x+\sigma )+s(x-\sigma )-2s(x)}{\sigma ^{2}}\right\} \label{deq:differentlimitdiffusion} 
\end{eqnarray}
 Now letting \( \tau _{0},\sigma \rightarrow 0 \) such that \( \sigma ^{2}/\tau _{0}\rightarrow 2D \)
we obtain (\( \Phi \equiv \Phi (x) \), \( p\equiv p(x,t) \)) \begin{eqnarray}
 \frac{\partial p}{\partial t} & = & \mathcal{L}_{G}\, p\label{eq:lgdef} \\
 & = & e^{-\beta \Phi /2}\left( \frac{\partial }{\partial x}\right) ^{2}e^{\beta \Phi /2}\, p-p\, e^{\beta \Phi /2}\left( \frac{\partial }{\partial x}\right) ^{2}e^{-\beta \Phi /2},\label{eq:lgrepres} 
\end{eqnarray}
 i.e. equation (\ref{FPEvar}). Therfore, the dynamics of a particle on a fully extended polymer is governed by ordinary diffusion.

Let us now return to a flexible polymer performing conformational
changes. In this case the geometrical factor in the rate (\ref{eq:rate}),
\( f(x-y) \) follows a power-law~\cite{Sok0}. Thence, we investigate
the case of long-tailed transition rates. 
Let  \begin{equation}
\label{eq:fdefepsilontail}
f(x)=\left\{ \begin{array}{ll}
C_{\mu }|x|^{-(1+\mu )} & \quad |x|>\varepsilon >0\\
0 & \quad \mathrm{otherwise}
\end{array}\right. 
\end{equation}
 with \( C_{\mu }= \Gamma (1+\mu )\sin (\pi \mu /2)/\pi  \) and
\( 0<\mu \leq 2 \).  Here \( \varepsilon \) is the minimal jump length, which
substitutes a fixed jump length \( \sigma \) of our previous example.
Note that we do not require $f(x)$ in (\ref{eq:fdefepsilontail}) to be
normalized to unity. Any prefactor independent of $\varepsilon$ 
occurs in both terms in the master equation (\ref{eq:mastergleichung}) 
and can be absorbed as a time constant. 
The particular choice of $C_\mu$ will become clear below.
Inserting (\ref{eq:fdefepsilontail}) 
into (\ref{eq:rate}) gives \begin{eqnarray}
\frac{\partial p(x,t)}{\partial t} & = & C_{\mu }\left\{ e^{-\beta \Phi (x)/2}\int _{|x-y|>\varepsilon }dy\, \frac{e^{\beta \Phi (y)/2}p(y,t)}{|x-y|^{1+\mu }}\right. \nonumber \\
 &  & -\left. e^{\beta \Phi (x)/2}p(x,t)\int _{|x-y|>\varepsilon }dy\, \frac{e^{-\beta \Phi (y)/2}}{|x-y|^{1+\mu }}\right\} \label{deq:masterepsitail} 
\end{eqnarray}
 In the limit \( \varepsilon \rightarrow 0 \) (\ref{deq:masterepsitail})
becomes (for notational ease we set \( s(x)\equiv e^{-\beta \Phi (x)/2} \),
\( q(x,t)=p(x,t)/s(x) \)) 
\begin{eqnarray}
\frac{\partial p(x,t) }{\partial t}  & = & \lim _{\varepsilon \rightarrow 0}C_{\mu }\left\{ s(x)\int _{|x-y|>\varepsilon }dy\, \frac{q(y,t)}{|x-y|^{1+\mu }}-\right. \label{deq:epsilon1} \\
 &  & -\left. q(x,t)\int _{|x-y|>\varepsilon }dy\, \frac{s(y)}{|x-y|^{1+\mu }}\right\} \nonumber \\
 & = & \lim _{\varepsilon \rightarrow 0}C_{\mu }\left\{ s(x)\int _{|x-y|>\varepsilon }dy\, \frac{q(y,t)-q(x,t)}{|x-y|^{1+\mu }}\right. \label{deq:epsilon2} \\
 &  & -\left. q(x,t)\int _{|x-y|>\varepsilon }dy\, \frac{s(y)-s(x)}{|x-y|^{1+\mu }}\right\} \nonumber \\
 & = & C_{\mu }\left\{ s(x)\int dy\, \frac{q(y,t)-q(x,t)}{|x-y|^{1+\mu }}\right. \label{deq:epsilon3} \\
 &  & -\left. q(x,t)\int dy\, \frac{s(y)-s(x)}{|x-y|^{1+\mu }}\right\} .\nonumber \label{deq:epsilon4} 
\end{eqnarray}
Note, that in the limit the resulting rates $w(x|y)$ are not normalizable. Therefore,
as mentioned earlier, the process is not a pure jump process in this limit.
Implicitely, in the limit carried out above, not only the minimal step size
$\varepsilon$ but also the typical waiting time $\tau_w(y)=(\int dx w(x|y))^{-1}$ at position $y$
vanishes, because of the singular nature of the rate $w(x|y)$ in this limit.
However, the limit can be carried out since the resultant singularities cancel
and eq.(\ref{deq:epsilon3}) can be interpreted consistently. 

The integrals appearing in (\ref{deq:epsilon3}) are symmetric fractional
generalizations of the ordinary Laplacian (see Appendix 1) \begin{equation}
\label{deq:nablamu}
\triangle ^{\frac{\mu }{2}}f(x)=\frac{\Gamma (1+\mu )\sin (\pi \mu /2)}{\pi }\int dy\, \frac{f(y)-f(x)}{|x-y|^{1+\mu }},
\end{equation}
 and equation (\ref{deq:epsilon3}) can be recast into a more concise
form: \begin{eqnarray}
\frac{\partial p}{\partial t}  & = & e^{-\beta \Phi /2}\, \triangle ^{\frac{\mu }{2}}\, e^{\beta \Phi /2}\, p-p\, e^{\beta \Phi /2}\, \triangle ^{\frac{\mu }{2}}\, e^{-\beta \Phi /2}\label{deq:lgmu} \\
 & \equiv  & \mathcal{L}_{G,\mu }\, p.\label{deq:lgmu2} 
\end{eqnarray}
 Again, the similarity to the underlying master-equation is obvious.
Trivially, detailed balance is retained in the limit and the Boltzmann
distribution is the stationary state of (\ref{deq:lgmu}). Note
that for \( \mu =2 \) we recover system (\ref{eq:lgdef}), i.e.
\begin{equation}
\label{deq:firstequivalence}
\mathcal{L}_{G,2}=\mathcal{L}_{FP}.
\end{equation}
 However, in the range \( 0<\mu <2 \) the operator \( \mathcal{L}_{G,\mu } \)
does \emph{not} portray a situation equivalent to any of the equations
discussed above and does not correspond to any Langevin equation of
the type \begin{equation}
\label{deq:generallangevin}
dX=F(X)dt+dL_{\mu }(t)
\end{equation}
 in which the long-tailed influence is incorporated into the thermal
fluctuations by a symmetric L\'{e}vy stable process \( L_{\mu }(t) \).
In~\cite{brocki2} the difference between the fractional
 Fokker-Planck equation 
corresponding to  L\'{e}vy-stable additive noise and
the topological generalization is investigated in detail.

\section{Comparison of systems in a double well potential}
\label{compa}

\begin{figure}
{\centering \resizebox*{12cm}{!}{\includegraphics{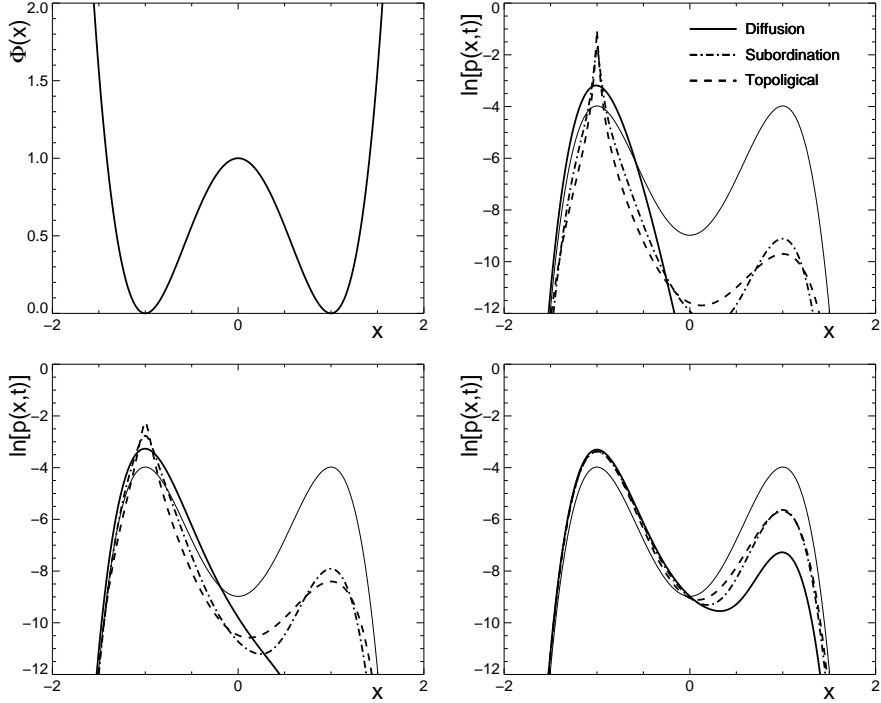}} \par}

\caption{\label{fig:pxt}Evolution of densities \protect\( p(x,t)\protect \)
in the double well potential \protect\( \Phi (x)\protect \) (upper
left) at three different times: \protect\( t=0.01\protect \) (upper
right), \protect\( 0.1\protect \) (lower left) and \protect\( 1.0\protect \)
(lower right) for \protect\( \beta =5.0\protect \) and \protect\( \mu =1\protect \)
with initial condition \protect\( p(x,0)=\delta (x+1)\protect \)
in relation to the stationary state (thin solid line). The diffusion
process equilibrates quickly within the left potential well and does
not pass the barrier for short times. In contrast, the superdiffusive
processes transfer probability even for small times and remain peaked
at \protect\( x=-1\protect \) for much longer.}
\end{figure}

We proceed by showing that the two generalizations of the FPE leading
to Boltzmann statistics in equilibrium (one based on the fractal time,
another on the fractal space approach) differ strongly in their predictions
concerning relaxation in the same kind of potential. For our comparison
we have chosen a classical Kramers situation, i.e. diffusion in a
double-well potential. As a generic double well we chose\begin{equation}
\label{deq:pot}
\Phi (x)=x^{4}-2x^{2}+1
\end{equation}
depicted in the upper left panel of figure \ref{fig:pxt}. Initially a particle
is placed at the potential minimum at \( x=-1 \), \( p(x,0)=\delta (x+1) \).
The solutions \( p(x,t) \) are depicted at three different times
(\( t=0.03,\, 0.1,\, 1.0 \)) in figure \ref{fig:pxt} for a value
\( \beta =5.0 \) and an exponent \( \mu =1 \). The fractional diffusion
equations were numerically solved by first mapping them onto the corresponding
hermitian Hamiltonian \( \mathcal{H} \) and subsequent spectral decomposition
\( \mathcal{H}=\sum _{n}\lambda _{n}\mathcal{P}_{n} \) where the
\( \lambda _{n} \) and \( \mathcal{P}_{n} \) are eigenvalues and
projectors onto the corresponding eigenspaces. Solutions \( p(x,t) \)
were then expressed in term of these quantities.

\begin{figure}
{\centering \resizebox*{1\textwidth}{!}{\includegraphics{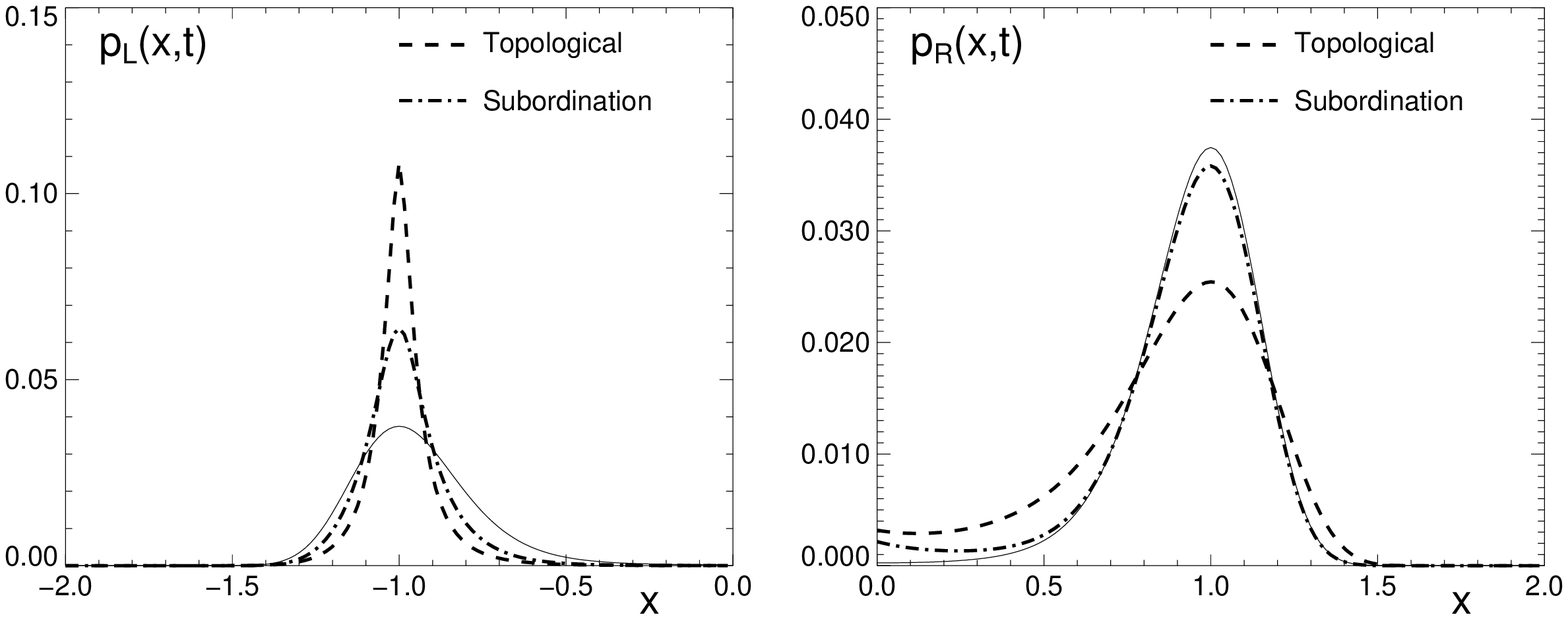}} \par}

\caption{\label{fig:pxt2}The shape of \protect\( p_{L}(x,t)\protect \) (left)
and \protect\( p_{R}(x,t)\protect \) (right) for subordinative and
topological superdiffusion at \protect\( t=0.1\protect \), \protect\( \beta =5.0\protect \)
and exponent \protect\( \mu =1\protect \). Note the difference in scale on the left and right, respectively. Despite the fact that
these processes are identical if \protect\( \Phi =0\protect \), they
respond quite differently if a potential is imposed.}
\end{figure}

Let us first note that the time-scales of the relaxation processes
for the case of normal diffusion and for L\'{e}vy-processes differ
vastly. The typical times of the concentration equilibration (being
the inverse of the largest nonzero eigenvalue) are: \( \tau _{C}=20.36 \)
for the diffusion process (\( 2 \)), \( \tau _{C}=4.51 \) for the
subordination case (\( \mu =1 \)) and \( \tau _{C}=3.93 \) for the
case of long jumps with the same value of \( \mu =1 \). A much greater
value of \( \tau _{C} \) for the ordinary diffusion process is not
surprising. In a short time it equilibrates within the left potential
well. However, due to the continuity of its sample paths, the diffusion
process accumulates probability in the right potential well only for
long times. In comparison, the subordination process as well as the
process governed by (\ref{deq:lgmu2}) can pass the potential maximum
at \( x=0 \) even for short times, a direct consequence of the possibility
of initiating a long jump from left to right. The possibility to transmit
probability through the potential barrier is payed off by a slower
equilibration within the left potential well in which \( p(x,t) \)
for both, subordination and topological superdiffusion stay sharply
peaked around \( x=-1 \). 
This cusp-like shape in the left potential well is
reminiscent of the classical Scher-Montroll picture of CTRW. 
Despite their qualitative similarities
in the overall shape of \( p(x,t) \) subordination and topological
superdiffusion are quite distinct on close inspection as is shown
in figure \ref{fig:pxt2}. 

In order to avoid huge difference in scales, we plot in figure 4 the reduced densities
in the left and in the right potential wells, i.e. the densities which are 
normalized to unity on the given sub-interval and for the given time:
\begin{eqnarray}
p_{L}(x,t) & = & \frac{p(x,t)}{\int _{-\infty }^{0}dy\, p(y,t)}\label{deq:pl} \\
p_{R}(x,t) & = & \frac{p(x,t)}{\int _{0}^{\infty }dy\, p(y,t)}\label{deq:pr} 
\end{eqnarray}
Clearly, the shapes differ strongly on both sides of the potential.

\begin{figure}
{\centering \resizebox*{12cm}{!}{\includegraphics{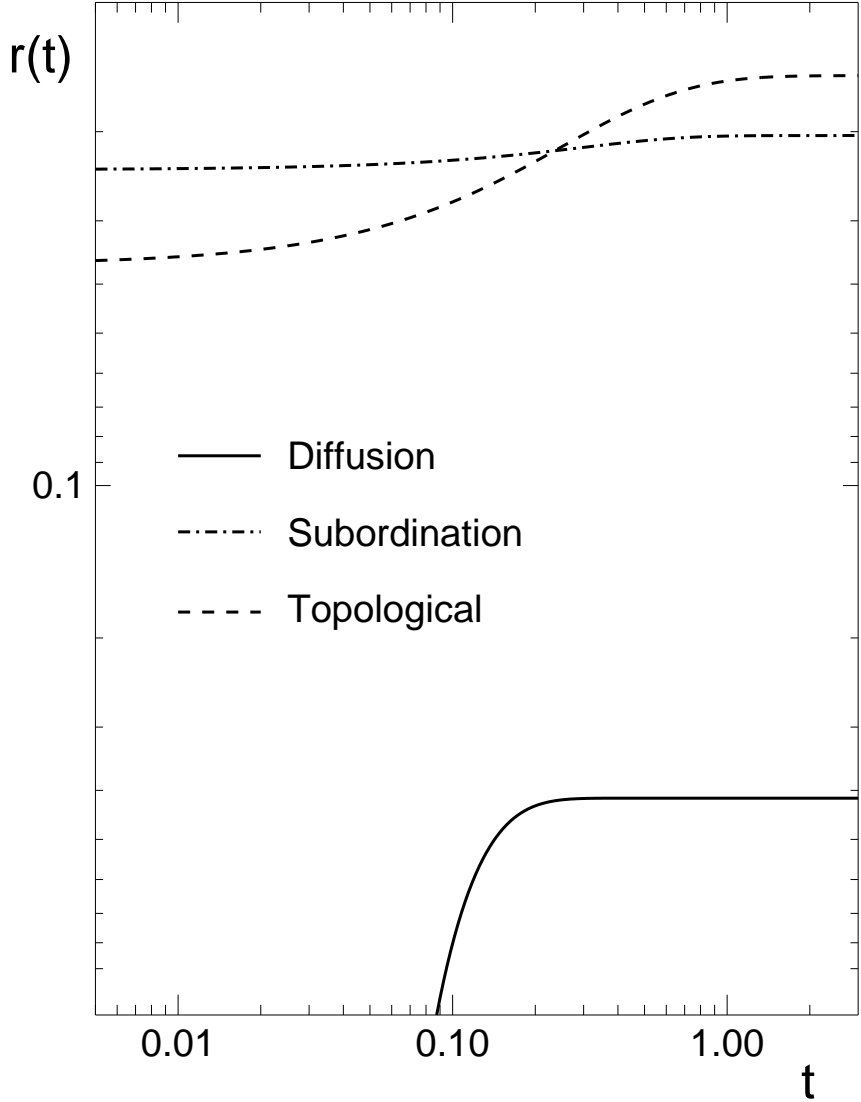}} \par}

\caption{\label{fig:rate}The time dependent rate \protect\( r(t)\protect \)
for ordinary diffusion, subordination and topological superdiffusion
in the double well. Parameters are \protect\( \beta =5.0\protect \),
\protect\( \mu =1\protect \).}
\end{figure}

In order to reveal the differences in barrier penetration it is instructive
to consider the time dependence of the total probability in the right potential
well
\begin{equation}
\label{deq:Q}
Q(t)=\, \int _{0}^{\infty }dx\, p(x,t),
\end{equation}
and the rate \( r(t) \) defined by\begin{equation}
\label{deq:r}
r(t)=-\frac{1}{2}\frac{d\ln Q(t)}{dt}.
\end{equation}
Asymptotically, as \( t\rightarrow \infty  \) the processes investigated
here evolve according to \begin{equation}
\label{deq:asymp}
p(x,t)\approx p_{s}(x)-\left( p_{s}(x)-p_{0}(x)\right) e^{\lambda _{1}t}\quad \lambda _{1}<0.
\end{equation}
where \( p_{s}(x) \) and \( p_{0}(x) \) denote the stationary and
initial states of the system, respectively. Therfore, \( r(t)\rightarrow -\lambda _{1} \)
as \( t\rightarrow \infty  \), the rate approaches the inverse of
the relaxation time constant of the process. In other words, the function
\( r(t) \) measures how fast the probability transfer across the
barrier approaches the quasi-equilibrium value. For the same parameters
as in figures \ref{fig:pxt}, \ref{fig:pxt2} the rate \( r(t) \)
is depicted in figure \ref{fig:rate}. The ordinary diffusion 
rate is initially zero and increases as soon as the process equilibrates
within the left potential well. It then levels off to a relatively small constant value
proportional to the probability flux across the barrier. In contrast,
for  the superdiffusive systems the
rate to transfer probability over the barrier is finite even for very small times. 
For the latter
\( r(t) \) increases more slowly than in ordinary diffusion. However,
the asymptotic rate is higher. Note that the subordination process
penetrates the barrier at relatively high but almost constant rate
for the entire time, whereas for topological superdiffusion the rate
is initially smaller but increases for large times beyond the subordination
rate. 

\section{Conclusions}

We discussed different generalizations of a Fokker-Planck equation
for the case of L\'{e}vy processes in external force fields. We show
that such generalization can proceed along different lines (assuming
strong temperature or mobility fluctuations or allowing for long jumps),
which lead to different forms of the generalized operator. We confined
ourselves to situations in which the corresponding equations are thermodynamically
sound and their solutions approach 
Boltzmann equilibrium.
The different relaxation behaviors of L\'{e}vy-processes
were investigated and compared for the paradigmatic case 
of a double-well potential and deviations from normal
diffusion were clarified.

\section{Acknowledgments}

The authors acknowledge useful discussions with J. Klafter and A.
Chechkin and W. Noyes. 
The financial support of the Fonds der Chemischen Industrie
is gratefully acknowledged.

\appendix

\section{Fractional Differentiation}

The operators involved in fractional differentiation are generically
non-local. Thus, depending on the particular choice of boundary values
a variety of generalizations exists, all of which suit a specific
need in physical applications. 
Depending on the specific problem at
hand different representations of fractional differential operators
may be chosen. Here we concentrate on a fractional generalization,
\( \triangle ^{\mu /2} \), \( 0<\mu \leq 2 \) of the ordinary Laplace
operator \( \triangle  \) generically encountered in superdiffusive
systems. The reader is refered to~\cite{oldham} for
additional information on fractional calculus in general.

\subsection{Definitions}

Incidentally, the most natural way to introduce fractional differential
operators is based on the generalization of the \( n \)-fold iterated
integral of a function \( f(x) \)\begin{equation}
\label{apxeq:nfoldint}
_{a}I_{x}^{n}\, f(x)\equiv \frac{1}{(n-1)!}\int _{a}^{x}dy\, f(y)(x-y)^{n-1}=\frac{1}{\Gamma (n)}\int _{a}^{x}dy\, f(y)(x-y)^{n-1}
\end{equation}
to arbitrary fractional order \( \alpha >0 \), \begin{eqnarray}
_{a}I_{x}^{\alpha }\, f(x) & \equiv  & \frac{1}{\Gamma (\alpha )}\int _{a}^{x}dy\, f(y)(x-y)^{\alpha -1}\label{apxeq:nfoldintalpha} \\
_{x}I_{a}^{\alpha }\, f(x) & \equiv  & \frac{1}{\Gamma (\alpha )}\int _{x}^{a}dy\, f(y)(y-x)^{\alpha -1},\label{apxeq:nfoldintalphaleft} 
\end{eqnarray}
 where we distinguish between the right- and left-handed fractional
integrals \( _{a}I_{x}^{\alpha } \) and \( _{x}I_{a}^{\alpha } \),
respectively. A fractional derivative of order \( \mu  \) which is
frequently employed in physical applications is given by the \( n^{\mathrm{th}} \)
ordinary derivative of the fractional integral of order \( 0<n-\mu \leq 1 \),
where \( n=[\mu ]+1 \). In other words,
\begin{eqnarray}
_{a}D_{x}^{\mu }f(x) & \equiv  & \frac{d^{n}}{dx^{n}}{_{a}I_{x}^{n-\mu }}\label{apxeq:fracderiv1} \\
_{x}D_{a}^{\mu }f(x) & \equiv  & \frac{d^{n}}{dx^{n}}{_{x}I_{a}^{n-\mu }}.\label{apxeq:fracderiv2} 
\end{eqnarray}
 Note that since the distinction between left- and right-handed fractional
integrations is passed on to fractional differentiation, fractional
derivatives are not symmetric with respect to interchanging their
subscripts. The definitions (\ref{apxeq:fracderiv1},\ref{apxeq:fracderiv2})
imply that \[
_{a}D_{x}^{m}=\frac{d^{m}}{dx^{m}}=(-1)^{m}{_{x}D_{a}^{m}}\quad m=0,1,2,...\]
 That is, fractional differentiation is proportional to ordinary differentiation
for integer values of the exponent. \footnote{%
To recover symmetry for integer values of the exponent \( m \) one
frequently encounters an alternative definition of the left-handed
fractional derivative, namely \( _{x}D_{a}^{\mu }f(x)\equiv (-1)^{n}\frac{d^{n}}{dx^{n}}{_{x}I_{a}^{n-\mu }} \)
which includes the pre-factor \( (-1)^{n} \) in the definition.} 
By sequential partial integration (\ref{apxeq:fracderiv1}) can
be recast into\begin{equation}
\label{apxeq:fracderivcommute1}
_{a}D_{x}^{\mu }f(x)=\sum ^{n-1}_{k=0}\frac{f^{(k)}(a)}{\Gamma (1+k-\mu )}(x-a)^{k-\mu }+_{a}I_{x}^{n-\mu }\frac{d^{n}}{dx^{n}}
\end{equation}
 Therfore, the operations of fractional differentiation and integration
generally do not commute. However, if we let \( a\rightarrow \pm \infty  \)
and require that \( \lim _{a\rightarrow \pm \infty }f^{(k)}(a)<\infty  \)
(\ref{apxeq:fracderiv1}) and (\ref{apxeq:fracderiv2}) yield\emph{\begin{eqnarray}
_{-\infty }D_{x}^{\mu }f(x) & = & \frac{1}{\Gamma (n-\mu )}\int _{-\infty }^{x}dy\frac{f^{(n)}(y)}{(x-y)^{\mu +1-n}}\label{eq:dinfxdef} \\
_{x}D_{\infty }^{\mu }f(x) & = & \frac{1}{\Gamma (n-\mu )}\int _{x}^{\infty }dy\frac{f^{(n)}(y)}{(y-x)^{\mu +1-n}}.\label{eq:dxinfdef} 
\end{eqnarray}
} In fractional diffusion equations the symmetric fractional generalization
of the Laplace operator frequently appears. Its one dimensional variant
can be defined in terms of \( _{a}D_{x}^{\mu }f(x) \) and \( _{x}D_{a}^{\mu }f(x) \)
by
\begin{equation}
\label{eq:nablamu}
\triangle ^{\frac{\mu }{2}}\equiv -\frac{1}{2\cos (\pi \mu /2)}\left\{ _{-\infty }D_{x}^{\mu }+_{x}D_{\infty }^{\mu }\right\} \quad 0<\mu \leq 2.
\end{equation}
 The particular case of \( \triangle ^{\frac{1}{2}} \) requires some
care. Although the definition (\ref{eq:nablamu}) does not give a
useful result for \( \mu =1 \), since both numerator and denominator
vanish, the limit \( \mu \rightarrow 1 \) can be interpreted consistently
as will become clear below. Partial integration of (\ref{eq:dinfxdef},\ref{eq:dxinfdef})
and insertion into (\ref{eq:nablamu}) yields \begin{equation}
\label{apxeq:laplatz2}
\triangle ^{\frac{\mu }{2}}f(x)=-\frac{1}{2\, \Gamma (-\mu )\cos (\pi \mu /2)}\int _{-\infty }^{\infty }dy\frac{\left[ f(y)-f(x)\right] }{|x-y|^{1+\mu }}.
\end{equation}
 Since \( \mathrm{lim}_{\mu \rightarrow 1}\Gamma (-\mu )\cos (\pi \mu /2)=-\pi /2 \)
the operator \( \triangle ^{\frac{1}{2}} \) is well defined, i.e.
\begin{equation}
\label{apxeq:sqrtlaplatz}
\triangle ^{\frac{1}{2}}f(x)=\frac{1}{\pi }\int _{-\infty }^{\infty }dy\frac{\left[ f(y)-f(x)\right] }{(x-y)^{2}}.
\end{equation}
 Clearly, \( \triangle ^{\frac{1}{2}}\neq \nabla =d/dx \), the {}``square-root''
of the Laplacian is symmetric, whereas the first derivative is not\footnote{%
In the literature the symbol \( \nabla ^{\mu } \) is frequently used
instead of \( \triangle ^{\frac{\mu }{2}} \)with can lead to some
confusion if \( \mu =1 \).
}The operators \( _{-\infty }D_{x}^{\mu } \), \( _{x}D_{\infty }^{\mu } \)
and \( \triangle ^{\frac{\mu }{2}} \), possess a particularly simple
form when the equation involved are Fourier transformed. If we define
the Fourier transform of \( f(x) \) by
\begin{equation}
\label{eq:fourierdefapxx}
\mathcal{F}[f](k)\equiv \widetilde{f}(k)=\int dx\, e^{ikx}f(x)
\end{equation}
 the transformed equations (\ref{eq:dinfxdef}), (\ref{eq:dxinfdef})
and (\ref{apxeq:laplatz2}) read \begin{eqnarray}
\mathcal{F}[_{-\infty }D^{\mu }f](k) & = & (-ik)^{\mu }\widetilde{f}(k)\label{eq:fourierdinfx} \\
\mathcal{F}[D_{\infty }^{\mu }f](k) & = & (ik)^{\mu }\widetilde{f}(k)\label{eq:fourierdxinf} \\
\mathcal{F}[\triangle ^{\frac{\mu }{2}}f](k) & = & -|k|^{\mu }\widetilde{f}(k).\label{eq:fourierlaplatzmu} 
\end{eqnarray}
 Thus, the Fourier transform of fractional differentiation is equivalent
to a multiplication by \( |k|^{\mu } \). Because of this, the investigation
of equations containing operators such as \( \triangle ^{\frac{\mu }{2}} \) is
usually much easier when expressed in Fourier space. The simplicity
of eqs. (\ref{eq:fourierdinfx}), (\ref{eq:fourierdxinf}) and (\ref{eq:fourierlaplatzmu})
is often a reason to chose the latter as definitions of fractional
derivatives. On the other hand, the corresponding integral representations
(\ref{eq:dinfxdef}), (\ref{eq:dxinfdef}) and (\ref{apxeq:laplatz2})
offer a more direct picture of the characteristic properties of the
corresponding operator. For example, the Cauchy process obeys \( \partial _{t}\widetilde{p}(k,t)=-|k|\widetilde{p}(k,t) \)
in Fourier space, a simple equation indeed. By inverse Fourier transform
this gives \begin{eqnarray}
\partial _{t}p(x,t) & = & (\triangle ^{\frac{1}{2}}p)(x,t)\label{apxeq:cauchyx1} \\
 & = & \int dy\, [w(x|y)p(y,t)-w(y|x)p(x,t)]\label{apxeq:cauchyx2} \\
w(x|y) & = & \frac{1}{\pi (x-y)^{2}}\label{apxeq:cauchyrate} 
\end{eqnarray}
 In contrast to the Fourier representation it is obvious from eqs.
(\ref{apxeq:cauchyx2}) and (\ref{apxeq:cauchyrate}) that the fractional
operator \( \triangle ^{\frac{1}{2}} \) is non-local, involves singularities
and appearing in the example given above describes a jump process,
a fact not readily available in Fourier representation.

To complete the picture, let us mention that the form (\ref{eq:fourierlaplatzmu})
of the fractional Laplace operator can be generalized to arbitrary
dimensions \( n \). Let \( f(\mathbf{x}) \) be a function on \( \mathbb {R}^{n} \)
and \( \widetilde{f}(\mathbf{k}) \) the corresponding Fourier transform,
then \( \triangle ^{\frac{\mu }{2}} \) defined by \begin{equation}
\label{eq:nablanddef}
\mathcal{F}[\triangle ^{\frac{\mu }{2}}f](\mathbf{k})\equiv -|\mathbf{k}|^{\mu }\widetilde{f}(\mathbf{k})
\end{equation}
 yields upon Fourier inversion the integral representation \begin{equation}
\label{eq:nablaortnd}
\triangle ^{\frac{\mu }{2}}f(\mathbf{x})=-\frac{2^{\mu }\Gamma \left( \frac{\mu +n}{2}\right) }{\pi ^{n/2}\Gamma \left( -\frac{\mu }{2}\right) }\int _{-\infty }^{\infty }d^{n}y\frac{\left[ f(\mathbf{y})-f(\mathbf{x})\right] }{|\mathbf{x}-\mathbf{y}|^{\mu +n}}
\end{equation}

\end{document}